\documentclass[amsmath,amssymb, aps, prl,reprint,footinbib,floatfix]{revtex4-2}
\usepackage{float}
\usepackage{amsmath,amssymb}
\usepackage[english]{babel}
\usepackage{upgreek}
\usepackage{dsfont}
\usepackage{graphicx}
\usepackage{color}
\usepackage{bbold}
\usepackage{hyperref}

\begin{document}
\title{Interferometric control of nanorotor alignment}

\newcommand{\bsfm}{\mbox{\textbf{\textsf{c}}}}
\newcommand{\tcb}[1]{{#1}}

\author{Birthe Schrinski}
\affiliation{University of Duisburg-Essen, Faculty of Physics, Lotharstra\ss e 1, 47048 Duisburg, Germany}
\author{Benjamin A. Stickler}
\affiliation{University of Duisburg-Essen, Faculty of Physics, Lotharstra\ss e 1, 47048 Duisburg, Germany}
\author{Klaus Hornberger}
\affiliation{University of Duisburg-Essen, Faculty of Physics, Lotharstra\ss e 1, 47048 Duisburg, Germany}

\begin{abstract}
The intrinsically non-linear rotation dynamics of rigid bodies offer unprecedented ways to exploit their quantum motion. In this Letter we devise a rotational analog of Mach-Zehnder interferometry, which allows steering symmetric rotors from fully aligned to completely antialigned. The scheme uses a superposition of four distinct orientations, emerging at the eighth of the quantum revival time, whose interference can be controlled by a weak laser pulse. We develop a semiclassical model of the effect and demonstrate that it persists even in presence of imperfections and decoherence.
\end{abstract}
\maketitle

\textit{Introduction.---}It is a major aim in the field of optomechanics \cite{aspelmeyer2014,bowen2015} to control the motion of levitated nanoparticles at the quantum limit \cite{millen2020}, as required for fundamental tests and for precision sensing \cite{arndt2014a,millen2020b,rademacher2020,moore2020}. Levitated objects have been used in their classical state of motion to search for physics beyond the standard model \cite{rider2016,monteiro2020search} and to demonstrate force sensitivities at the zeptonewton level \cite{ranjit2016,hempston2017}. The recent achievement of  cooling the center-of-mass motion of a nanosphere to the ground state   \cite{delic2020,magrini2021realtime,tebbenjohanns2021quantum} now heralds a new era by mastering the quantum dynamics of internally warm solid objects composed of millions of atoms.

Levitated nanoparticles rotate, adding an intrinsically nonlinear twist to their center-of-mass dynamics.
To date, experiments with rotating particles still operate in the classical domain; they spin nanorotors with 
ultra-high precision \cite{kuhn2017b,reimann2018} and at ultra-high frequencies \cite{reimann2018,ahn2018,jin2020}, and demonstrate precession \cite{rashid2018,bang2020}, radiation-torque heating \cite{vanderlaan2020b}, and record-breaking torque sensitivities \cite{ahn2020}. First experimental implementations of rotational cooling \cite{delord_2020,bang2020,vanderlaan2020b} suggest that the quantum regime is within reach \cite{Stickler_2016cooling,seberson2019,tebbenjohanns2021optimal} and that even the trapped ground state of the full translational and rotational motion can be prepared \cite{schafer2021}.

Quantum rotations of molecules and nanoparticles provide unprecedented ways for quantum-enhanced torque sensing and for testing quantum physics \cite{stickler2021quantum}. However, it is still an open problem how to steer nanoparticle alignment in free flight, as needed for future sensing and metrology applications. In this Letter, we solve this problem by devising a Mach-Zehnder-type interference scheme operating in the curved and closed manifold of rigid body orientations. The ability to achieve such control over massive objects will enable unforeseen possibilities for orientation-resolved spectroscopy, rotation state-resolved collision and reaction studies, spatially resolved torque and rotation sensing, and quantum superposition tests \cite{koch2019,stickler2021quantum}.

The rotational analog of Mach-Zehnder interferometry is based on the phenomenon of orientational quantum revivals, which occur in free rotors as a direct consequence of angular momentum quantization. Specifically, the quadratic dependence of the rotational energies on the total angular momentum quantum number leads to the recurrence of the initial state \cite{Robinett_2004} at a characteristic revival time $T_{\rm rev}$, which may be much longer than the wave packet dispersion time. Such revivals have been predicted \cite{Seidemann_1999,machholm2001} and observed \cite{Poulsen_2004,goban2008,ghafur2009,de2009} in the alignment  of small molecules. Rotational quantum effects have also been proposed for  controlling \tcb{planar and linear} rotations \tcb{of molecules} \cite{Ivanov_2004}, for polarizability metrology \cite{Berglund2015}, for macroscopic quantum superposition tests \cite{Stickler_2018}, and for observing the quantum tennis-racket effect \cite{ma2020}.

\begin{figure}
\center
\includegraphics[width=0.46\textwidth]{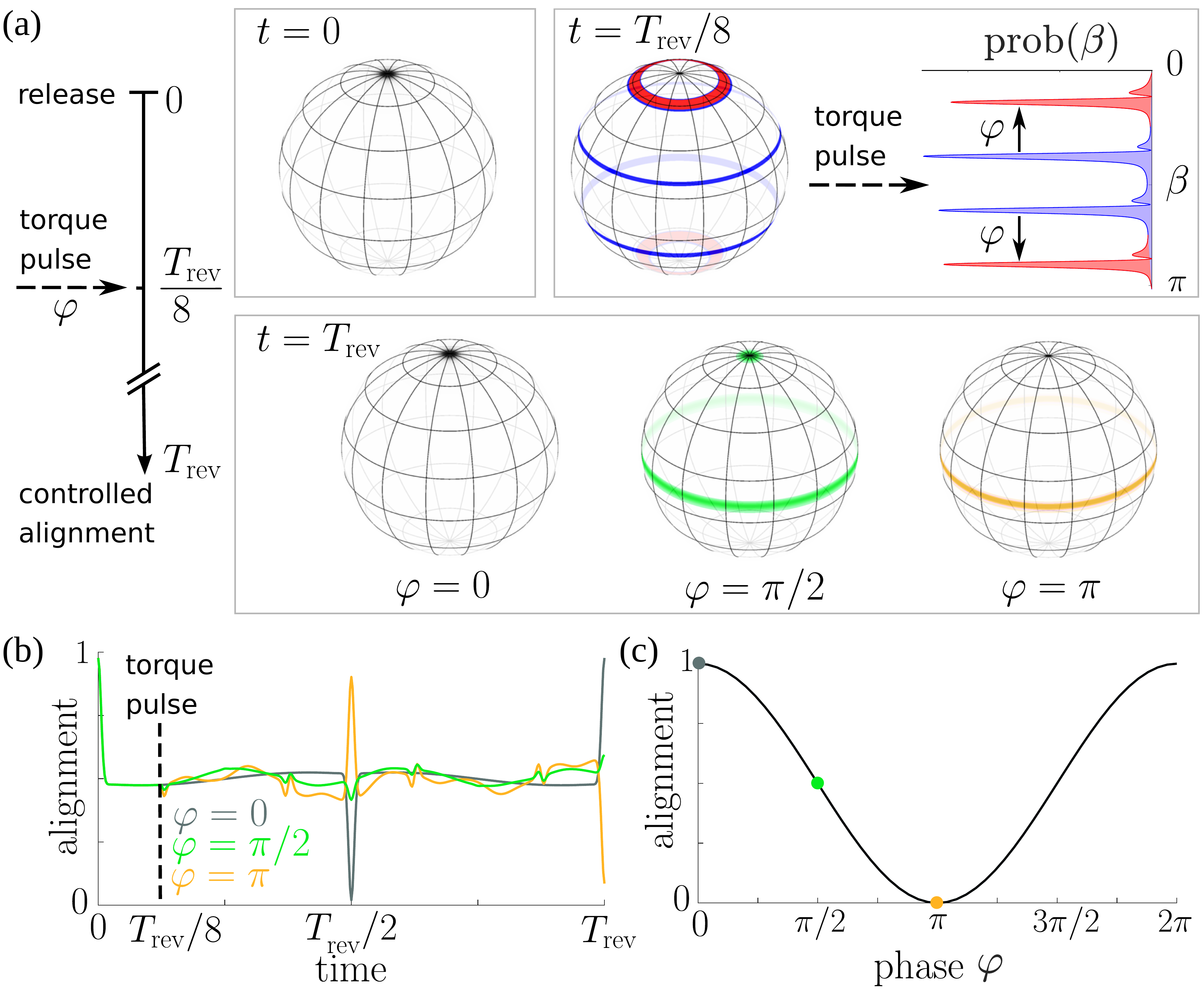}
\caption{(a) Orientational distribution of the rotor symmetry axis at different times $t$. An initially well-aligned state at the north pole (top left) recurs fractionally  as a superposition of wave packets with well-localized latitudes (top right) at  $t=T_\mathrm{rev}/8$.
A torque pulse imprints a relative phase $\varphi$  onto this superposition even if the interaction cannot distinguish between inverted orientations. Equal phases are indicated by equal colors in the corresponding polar angle distribution prob$(\beta)$. The torque pulse thus controls whether the state at $t=T_\mathrm{rev}$ is fully aligned ($\varphi=0$), completely antialigned ($\varphi=\pi$), or a balanced superposition thereof ($\varphi=\pi/2$).
(b) Time evolution of the corresponding alignment signal $\langle \cos^2\beta\rangle_t$ for $\varphi=0$ (gray), $\varphi=\pi/2$ (green) and $\varphi=\pi$ (orange). 
(c) Expected signal $\langle\cos^2\beta\rangle_{T_\mathrm{rev}}$ as a function of $\varphi$. 
}
\label{fig:1}
\end{figure}

Here we show how the 3D alignment of freely rotating nanoparticles can be controlled by applying a short torque pulse at $T_{\rm rev}/8$. It allows  tuning the alignment with respect to the initial orientation of the particle from fully aligned to completely antialigned, even in the generic case of torques not differing for inverted orientations.
This rotational interference scheme makes use of  superpositions of well-localized orientational wave packets emerging briefly at fractional revival times. \tcb{This generalizes coherent rotational control schemes for linear molecules \cite{Ivanov_2004} and condensed atom clouds \cite{Kialka_2020} to massive nanoparticles, exhibiting} genuine  non-commutative 3D rotations. We show that our scheme works even with millions of rotation states involved, for realistic particle asymmetries, and in presence of environmental decoherence.
	
\textit{Interference scheme.---}We first consider symmetric rotors to explain the rotational interference scheme in terms of a semiclassical eight-state model.
The free Hamiltonian \tcb{
$\mathsf{H}=\mathsf{{J}}^2/2I+(1/2I_c-1/2I)\mathsf{J}_c^2$} involves the square $\mathsf{{J}}^2$ of the angular momentum vector and its  body-fixed 
component $\mathsf{J}_c$, with $I_c$ and $I$  the moments of inertia around and orthogonal to the rotor symmetry axis ${\bf c}$ \cite{landaulifshitz1958}. In the free symmetric rotor eigenbasis $|jmk\rangle$ (with $|m|,|k|\leq j$
the quantum numbers for $\mathsf{J}_z$ and $\mathsf{J}_c$, and  $j\in\mathbb{N}_0$ the total angular momentum quantum number), the time evolution operator takes the form
$\mathsf{U}(t)=\sum_{jmk}e^{-i\pi[j(j+1)+(I/I_c-1)k^2]t/T_{\rm rev}}\vert jmk\rangle\langle jmk\vert$, with  $T_\mathrm{rev}=2\pi I/\hbar$.
Taking the direction of initial alignment as the space-fixed $z$-axis implies that 
the angular momentum representation of the aligned initial state $\rho_0$ is diagonal in $m$ and $k$. From this it readily follows that any such symmetric rotor state will fully recur, $\mathsf{U}(T_\mathrm{rev})\rho_0\mathsf{U}^\dagger(T_\mathrm{rev})=\rho_0$, as a direct consequence of angular momentum quantization.  

To characterize fractional revivals, consider a massive prolate rotor ($I_c\ll I$) in a  well-aligned pure initial state, $\rho_0=|\Psi_0\rangle\langle\Psi_0|$. Using Euler angles in the $z$-$y'$-$z''$-convention \cite{edmonds1957}, the angle  $\beta\in[0;\pi]$ between the $z$-axis and the rotor axis is then localized close to $\beta=0$, implying that the quantum numbers of $\mathsf{J}_z$ and $\mathsf{J}_c$ coincide, $\langle \alpha,\beta,\gamma\vert \Psi_0\rangle=\langle\beta\vert\psi_0\rangle
\exp[i k_0(\alpha+\gamma)]
/2\pi$. The total angular momentum quantum number $j$ will be distributed over a wide range of large values so that the Wigner d-matrix elements in $\langle \alpha\beta\gamma\vert jmk\rangle=(j+1/2)^{1/2}d_{mk}^j(\beta)\exp(im\alpha+ik\gamma)/2\pi$ can be replaced by their asymptotic expressions \cite{littlejohn2009} for $|m|,|k|\ll j$,
\begin{align}\label{eq:Wignerdsemi}
d_{mk}^j(\beta)\simeq\frac{\cos\left[(j+\frac{1}{2})\beta+(m-k)\frac{\pi}{2}-\frac{\pi}{4}\right]}{\sqrt{\frac{\pi}{2} (j+\frac{1}{2})\sin\beta}}\,.
\end{align}
We focus first on the time evolution $|\psi_t\rangle=\mathsf{U}_\beta |\psi_0\rangle$ for $m=k=0$ and consider finite occupations of $\mathsf{J}_z$, $\mathsf{J}_c$  later on.

The semiclassical propagator for the $\beta$-motion can then be decomposed as
\begin{align}
\begin{split}
\langle\beta\vert\mathsf{U}_\beta(t)\vert\beta_0\rangle\simeq\frac{u_c(\beta-\beta_0;t)+u_s(\beta+\beta_0;t)}{\sqrt{\sin\beta\sin\beta_0}}
\end{split}
\label{eq:scpropagator}
\end{align}
with
\begin{subequations}\label{eq:ucus}
\begin{align}
    u_c(\beta;t)&=\frac{1}{\pi}\sum_{j=0}^\infty e^{-i\pi j(j+1)t/T_\mathrm{rev}}\cos\Big[\Big(j+\frac{1}{2}\Big)\beta\Big],\label{eq:uc}\\
    u_s(\beta;t)&=\frac{1}{\pi}\sum_{j=0}^\infty e^{-i\pi j(j+1)t/T_\mathrm{rev}}\sin\Big[\Big(j+\frac{1}{2}\Big){\beta}\Big].\label{eq:us}
\end{align}
\end{subequations}
These contributions can be resummed at integer fractions of the revival time by using the expressions
\begin{subequations}
\begin{align}
    \sum_{j=0}^\infty\cos(j\theta)&=\pi\sum_{n=-\infty}^{\infty}\delta(\theta-2\pi n)+\frac{1}{2},\\
    \sum_{j=0}^\infty\sin(j\theta)&=\frac{1}{2}\mathcal{P}\cot\left(\frac{\theta}{2}\right).
\end{align}
\end{subequations}
with $\mathcal{P}$ the Cauchy principle value. They hold for smooth $2\pi$-periodic test functions and can be obtained from the Poisson summation formula.

Specifically, for $t=T_\mathrm{rev}/8$ the summands in (\ref{eq:ucus}) can be grouped into four sets of equal phase by splitting the summation index set into residue classes modulo 16. Using trigonometric addition theorems one finds
\begin{subequations}\label{eq:T8}
\begin{align}
\label{eq:ucsemi}
&u_c\left(\beta;\frac{T_\mathrm{rev}}{8}\right)=\frac{\sqrt{2}}{4}e^{-i 3 \pi/16}\sum_{n=0}^3  e^{-in(n+1)\pi/8}\nonumber\\
&\times\left\{\delta\left[\beta-(2n+1)\frac{\pi}{8}\right]+\delta\left[\beta+(2n+1)\frac{\pi}{8}\right]\right\},\\
\label{eq:ussemi}
&u_s\left(\beta;\frac{T_\mathrm{rev}}{8}\right)=\frac{1}{\pi}\sin\left(\frac{\beta}{2}\right)+\frac{2}{\pi}\sum_{n=0}^3e^{-in(n+1)\pi/8}\nonumber\\ 
&\times \mathcal{P}\cot(8{\beta})\left\{\cos\left[(2n+1){\beta}\right]-\cos\left[(15-2n){\beta}\right]\right\}.
\end{align}
\end{subequations}
The locations of the delta functions and Cauchy singularities in these expressions imply that a state initially localized at $\beta=0$ will be promoted, at $T_\mathrm{rev}/8$, into a superposition of narrow wave packets localized at $\beta=\pi/8$, $3\pi/8$, $5\pi/8$, $7\pi/8$ due to the constructive interference of all angular momentum states.

The state at $T_\mathrm{rev}/4$ is obtained by 
applying the propagator (\ref{eq:T8}) twice, or alternatively by resumming the (\ref{eq:ucus}) in residue classes modulo 8.
One finds that the initial state is promoted into 
a superposition of two wave packets localized at $\beta=\pi/4$, $3\pi/4$. Similarly, after half of the revival time the initial wave packet reappears localized at $\beta=\pi/2$, implying a perfectly antialigned rotor.

It thus follows from the composition property of 
$\mathsf{U}_\beta(t)$
that the states $\vert\psi_\ell\rangle=\mathsf{U}_\beta(\ell T_{\rm rev}/8)\vert \psi_0\rangle$ 
are composed of well-localized wave packets. Denoting by $|\xi_n\rangle$ the wave packet centered at $\beta=n\pi/8$, the coefficients in $\vert\psi_\ell\rangle=e^{i\nu_\ell}\sum_{n=1}^7M_{\ell n}\vert \xi_n\rangle$ can be read off from the semiclassical propagators 
($\ell=1,\ldots,7$). They are given by the unitary matrix
\begin{align}
\begin{split}
    M_{\ell n}&=\frac{1}{2}
    \begin{pmatrix}
    1 & 0 & 1 & 0 & 1& 0& 1 \\
     0 & \sqrt{2} & 0 & 0 & 0&\sqrt{2}& 0 \\
   1 & 0 &i & 0 & -i& 0& -1 \\
     0 & 0 & 0 & 2 & 0& 0& 0 \\
1 & 0 & -1 & 0 & -1& 0& 1 \\
     0 & \sqrt{2} & 0 & 0 & 0& -\sqrt{2}& 0 \\
     1 & 0 & -i & 0 & i& 0& -1 
 \end{pmatrix}
 \end{split}
\end{align}
and $\nu_\ell=(0,0,-\pi/8,0,\pi/2,\pi/4,3\pi/8,0)$.

The states $|\psi_{0}\rangle, \ldots, |\psi_7\rangle$ thus span an eight-dimensional subspace. It is well suited for interferometric control of the rotor alignment since phase differences can be imprinted by applying a torque for a brief duration at one or more of the fractional times $\ell T_{\rm rev}/8$. 

Torque pulses can be realised with an optical pulse polarized in $z$-direction \cite{koch2019}. The interaction energy is then proportional to $\cos^2\beta$ for particles characterised by an optical anisotropy axis, rendering their optical response invariant under the inversion of the orientation $\beta\to\pi-\beta$. The corresponding phase operator takes the form
\begin{align}
\upphi=\exp\left(i\sqrt{2}\varphi\cos^2\upbeta\right).
\label{eq:phaseoperator}
\end{align} 
It is diagonal in the angle operator $\upbeta$, with the phase $\varphi$ determined by the electric field $E(t)$ and the particle polarizability anisotropy $\Delta \alpha$, $\varphi = \Delta \alpha \int dt |E(t)|^2/2\sqrt{2}\hbar$.

To control the nanoparticle alignment one applies (\ref{eq:phaseoperator}) after the rotor has evolved freely for the time $T_\mathrm{rev}/8$. 
Abbreviating  $\tilde{\mathsf{U}}\equiv\mathsf{U}_\beta(T_{\rm rev}/8)$, this yields the state
\begin{align}
\begin{split}
\upphi\tilde{\mathsf{U}}\vert \psi_0\rangle=&\frac{\vert \xi_1\rangle+\vert\xi_7\rangle}{2}
+e^{-i\varphi}\frac{\vert\xi_3\rangle+\vert\xi_5\rangle}{2},
\end{split}
\end{align}
where we dropped a global phase. It has a relative phase $\varphi$ between wave packets located at arctic and tropic latitudes, see Fig.~\ref{fig:1}(a).
Quantum interference during the free evolution until the revival time transforms this into a superposition of the well-aligned initial state $|\psi_0\rangle$ and the  antialigned state $|\xi_4\rangle$,
\begin{align}
\begin{split}\label{eq:alignmentcontrol}
\tilde{\mathsf{U}}^7\upphi\tilde{\mathsf{U}}\vert \psi_0\rangle=\cos\left(\frac{\varphi}{2}\right)\vert \psi_0\rangle+\sin\left(\frac{\varphi}{2}\right)\vert \xi_4\rangle.
\end{split}
\end{align}
By tuning the phase $\varphi$ one can thus control the alignment at the revival time.

This is illustrated in Fig.~\ref{fig:1}, for an initial state with $|\langle j|\psi_0\rangle|^2\propto\exp(-j^2/800)$. 
It shows that alignment control is facilitated already at $T_{\rm ev}/2$ with the effects of $\varphi=0,\pi$  swapped.
Note that the revivals occur only for a short period of time, given by the initial alignment decay,
as determined by the angular momentum distribution.

For finite values of $m$ and $k$, it follows from (\ref{eq:Wignerdsemi}) that $u_{\rm c}(\beta,t)$ remains semiclassically unaffected as long as $|m|,|k|\ll j$, while  $u_{\rm s}(\beta,t)$ only acquires the additional sign $(-)^{m+k}$. Given that the initial state is well aligned, so that $m= k$, the eight-state model and its prediction (\ref{eq:alignmentcontrol}) therefore remain valid for weakly occupied $\mathsf{J}_z$, $\mathsf{J}_c$.

The numerical simulation of realistic particle states
requires matrix elements of the phase operator \eqref{eq:phaseoperator} in the angular momentum basis. Since their exact computation gets numerically intractable for large $j$, one can resort to the semiclassical approximation
\begin{widetext}
\begin{align}\label{eq:me}
\langle jmk\vert \upphi\vert j'm'k'\rangle= \delta_{mm'}\delta_{kk'}e^{-i\pi\vert j-j'\vert/4} \left[1 +i\sqrt{2\xi}\frac{32 k^2 m^2}{(j + j'+1)^4}\frac{d}{d\xi}\frac{1}{\sqrt{\xi}}\right]e^{-i A_{j+j'+1}^{km}/\xi}
J_\frac{\vert j-j'\vert}{2}\left(\frac{A_{j+j'+1}^{km}}{\xi}\right)\bigg\vert_{\xi=1/\varphi}
\end{align}
\end{widetext}
with $A_{J}^{km}=(1-4k^2/J^2)(1-4m^2/J^2)/\sqrt{2}$.
It can be obtained from the Bohr-Sommerfeld quantization of the associated action-angle variables \cite{child} in leading order of $mk/j^2$, and will be used below.

\textit{Alignment control of realistic nanoparticles.---}%
In practice, the initial state will not be perfectly aligned,  
the rotor will not be completely symmetric,
and the quantum dynamics will not be fully coherent. We discuss these imperfections in turn, showing that interferometric alignment control of realistic nanoparticles can still be expected in their presence.

For concreteness, consider an ellipsoidally shaped silicon nanorod with principal diameters
of 5.5\,nm and 50\,nm,
corresponding to a mass of 
$1.1\times 10^6$\,amu and a revival time of $T_\mathrm{rev}=14$ ms. 
After coherent scattering cooling close to the trapped ro-translational ground state in an elliptically polarized tweezer \cite{schafer2021}, one adiabatically changes to linear polarization such that the intrinsic rotation around the symmetry axis is released with only weakly occupied $\mathsf{J}_c$.
The initial state may then be described by a mixture of
$\langle\alpha,\beta,\gamma\vert\Psi_0\rangle \propto
\exp(-\sin^2\beta/4\sigma_\beta^2+i {k_0} (\alpha+\gamma))$
with $\sigma_\beta$ and $k_0$ determined by the cooling setup and the adiabatic release. Based on \cite{schafer2021} we assume 
$\sigma_\beta=3.1\times 10^{-3}~\mathrm{rad}$ and $k_0\in\mathbb{Z}$ distributed as a Gaussian with $\sigma_k <6$. 

Once the trapping laser is switched off, the particle falls freely, is illuminated by the weak phase pulse at $t=T_\mathrm{rev}/8$ and by a strong readout pulse around $T_\mathrm{rev}$, before it is recaptured and recooled. The torque needed to imprint the phase $\varphi$ must be applied for a time $\tau$ much shorter than the rotational dispersion time. A pulse with constant amplitude yields  $\varphi=\Delta\alpha\vert E_0\vert^2\tau/\sqrt{2}4\hbar$, where $\vert E_0\vert^2=4P/(\pi w_0^2\epsilon_0 c)$ depends on the  power $P$ and waist $w_0$ of the laser beam.

\begin{figure}
\center
\includegraphics[width=0.46\textwidth]{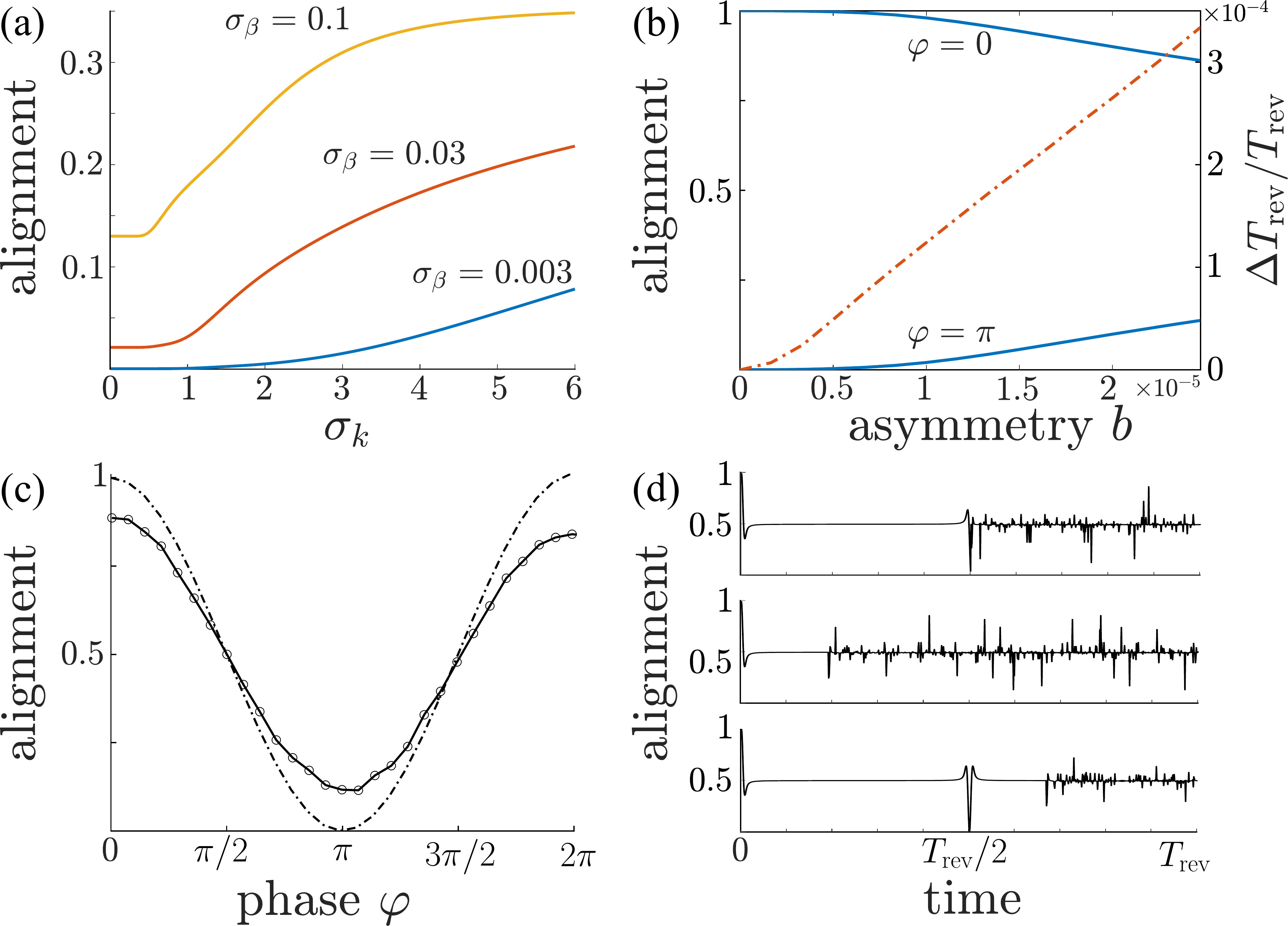}
\caption{(a) Antialignment effect at $\varphi = \pi$ for realistic initial states with angular spreads $\sigma_\beta=0.1$, $\sigma_\beta=0.03$, and $\sigma_\beta=0.003$ (top to bottom), as a function of the intrinsic angular momentum width $\sigma_k$. 
For perfectly aligned initial states the signal vanishes identically (\ref{eq:alignmentcontrol}). 
This interference effect gets impaired for realistic values of $\sigma_\beta$ and $\sigma_k$, but is still observable.
(b) Effect of nanoparticle asymmetry on the alignment for $\varphi=0$ and $\varphi=\pi$. The dash-dotted line shows how the asymmetry increases the revival time (orange, right scale). 
(c)  Collisional decoherence at a nitrogen gas pressure of $5\times 10^{-9}\,$ mbar reduces the achievable alignment control (\tcb{solid} line) compared to the case of perfect vacuum (\tcb{dash-dotted} line). (d) Alignment associated with three exemplary Monte-Carlo trajectories.
(The parameters used for all panels are given in the text, unless specified otherwise.)}
\label{fig:2}
\end{figure}

Figure \ref{fig:2}(a) shows how the alignment signal at $\varphi = \pi$ is affected by the uncertainty in the initial orientation and by finite values of the intrinsic angular momentum component. Both lead to a moderate reduction of the anti-alignment effect. This is due to the phase operator \eqref{eq:phaseoperator} no longer yielding definite relative phases for wave packets with angular dispersion, and due to the matrix elements (\ref{eq:me}) effecting an $m$- and $k$-dependent shift of the revival time. For $\varphi = 0$, in contrast, all symmetric rotors display a perfect alignment recurrence. 

The asymmetry of a general prolate top is characterized by the parameter $b=(I_a^{-1}-I_b^{-1})/(2I_c^{-1}-I_a^{-1}-I_b^{-1})$, which determines the approximate rotation energies in terms of the symmetric ones \cite{Davis_1961}.
Assuming one of the minor principal 
diameters to be 5.0\,nm corresponds to $|b|=2.3\times 10^{-5}$. For such values one may safely approximate the rotational eigenstates by those of a symmetric top, and take into account only the modified rotational energy spectrum.

Figure~\ref{fig:2}(b) shows the alignment signal at $\varphi = 0, \pi$ and the associated shift of the revival time  as a function of the asymmetry parameter for the initial state $|\Psi_0\rangle$ (with $\sigma_\beta=3\times 10^{-3}~\mathrm{rad}$ and $k_0 = 0$). One observes that shape asymmetries may lead to a noticeable reduction of the alignment effect, while only slightly shifting the revival time.
However, the asymmetry assumed above, which is well within the capabilities of present-day nanofabrication techniques, still yields a sizable alignment signal of $\langle\cos^2\beta\rangle=0.87$.

Collisions of the nanorotor with residual gas particles may effect a gradual diffusion of the nanoparticle angular momentum and thus provide a dominant source of decoherence \cite{Stickler_2016,Zhong_2016,Papendell_2017,stickler2018a}.   
Denoting the (operator-valued) orientation of the symmetry axis
by $\bsfm=(\cos\upalpha\sin\upbeta,\sin\upalpha\sin\upbeta,\cos\upbeta)^T$, and neglecting changes in the intrinsic rotation $\mathsf{J}_c$, the corresponding master equation reads as $\partial_t\rho=-i[\mathsf{H},\rho]/\hbar+\Gamma_{\rm gas}(\bsfm\cdot\rho\,\bsfm-\rho)$, with $\Gamma_{\rm gas}$ the collision rate \cite{Papendell_2017}.
It can be simulated by means of a Monte-Carlo unraveling \cite{Breuer,wiseman2009} in terms of the stochastic Schrödinger equation
\begin{align}\label{eq:sse}
\vert\mathrm{d}\Psi\rangle=\frac{1}{i\hbar}\mathsf{H}\mathrm{d}t+\sum_{\ell=1}^3\left(\frac{\mathsf{c}_\ell}{\sqrt{\langle \Psi| \mathsf{c}_\ell^2\vert\Psi\rangle}}-1\right)\vert\Psi\rangle\mathrm{d}N_\ell,
\end{align}
where the independent Poisson increments $\mathrm{d}N_\ell$ have the expectation values $\mathbb{E}[\mathrm{d}N_\ell]=\Gamma_{\rm gas}\langle\Psi|\mathsf{c}_\ell^2|\Psi\rangle\mathrm{d}t$. Here we used that $\sum_\ell{\sf c}_\ell^2={1}$, so that $\sum_\ell \mathbb{E}[{\rm d}N_\ell] = \Gamma_{\rm gas} {\rm d}t$ is state independent. 

Figure \ref{fig:2}(c) displays the impact of collisional decoherence on the interferometric alignment control for $\Gamma_{\rm gas}=20.7\,$Hz, corresponding to a nitrogen gas pressure of $5\times10^{-9}$\,mbar at room temperature. 
One observes that the interference signal gets slightly degraded for these realistic gas pressures, irrespective of the imprinted phase $\varphi$.
For illustration, Fig.~\ref{fig:2}(d) shows the  alignment dynamics of three sample trajectories from the ensemble described by (\ref{eq:sse}).

A further source of decoherence might be the Rayleigh scattering of laser photons from the phase pulse. However, this is negligible since a power of  $1.3$ mW suffices to imprint a  relative phase of $2\pi$ (assuming a pulse duration of $t_{\rm L}=100$\,ns, a wave length of $1550$\,nm, and a waist of $w_0=30\,\mu\mathrm{m}$), implying that only $1.7\times 10^{-13}$ photons scatter on average during the pulse.

\textit{Conclusions.---}%
The interferometric alignment scheme relies only on the free quantum dynamics  of symmetric rotors and on their polarization anisotropy. It can thus be employed for a wide range of particle species and sizes, 
even in absence of  internal spin \cite{delord_2020,delord2017a}, magnetization \cite{gieseler2020b}, or dipole moments \cite{afek2021control}.
The scheme will find applications whenever exquisite control is required  of the field-free alignment of single particles in vacuum \cite{koch2019}. 

We found that the interference effect can be represented in an effective eight-dimensional subspace spanned by superpositions of narrow orientational wave packets. More complex interference schemes are thus conceivable by applying several  pulses of light  at integer multiples of $T_{\rm rev}$/8, effecting unitary transitions between these states. They  could be used for preparing  orientational superposition states, for tests of quantum physics, for precision metrology of molecular properties, or even for processing quantum information  \cite{albert2020,grimsmo2020}. Beyond that, it is straightforward to use even smaller integer fractions of the revival time if interference in higher dimensional effective spaces is required.

\acknowledgments

This work was supported by the Deutsche Forschungsgemeinschaft (DFG—394398290).

\end{document}